\documentclass[runningheads]{llncs}
\usepackage[T1]{fontenc}
\usepackage{graphicx}
\usepackage{amsmath}
\usepackage{stmaryrd}
\usepackage{amsfonts}
\usepackage{color}

\DeclareMathOperator*{\argmin}{arg\,min}

\usepackage{bbm}
\usepackage{tabularray}
\usepackage{booktabs} 
\usepackage{siunitx} 
\usepackage[misc]{ifsym}

\usepackage[hidelinks]{hyperref}
\begin{document}
\title{Retinal IPA: \underline{I}terative Key\underline{P}oints \underline{A}lignment for Multimodal Retinal Imaging}
%
%
\author{Jiacheng Wang \inst{1}\Letter\and
Hao Li\inst{2} \and
Dewei Hu\inst{2} \and
Rui Xu\inst{3} \and
Xing Yao\inst{1} \and
Yuankai K.\ Tao \inst{3} \and
Ipek Oguz \inst{1,2,3}\Letter}

\institute{Department of Computer Science, Vanderbilt University \and 
Department of Electrical and Computer Engineering, Vanderbilt University \and
Department of Biomedical Engineering, Vanderbilt University \\
\email{\{jiacheng.wang.1,ipek.oguz\}@vanderbilt.edu}}

\authorrunning{J. Wang et al.}
%

%
\maketitle              
\begin{abstract}
We propose a novel framework for retinal feature point alignment, designed for learning cross-modality features to enhance matching and registration across multi-modality retinal images. Our model draws on the success of previous learning-based feature detection and description methods. To better leverage unlabeled data and constrain the model to reproduce relevant keypoints, we integrate a keypoint-based segmentation task. It is trained in a self-supervised manner by enforcing segmentation consistency between different augmentations of the same image. By incorporating a  keypoint augmented self-supervised layer, we achieve robust feature extraction across modalities. Extensive evaluation on two public datasets and one in-house dataset demonstrates significant improvements in performance for modality-agnostic retinal feature alignment. Our code and model weights are publicly available at \url{https://github.com/MedICL-VU/RetinaIPA}. 

\end{abstract}
\section{Introduction}


Retinal image alignment can be used to mosaic multiple images to create ultra-wide-field images \cite{wang2023novel} for a more comprehensive assessment of the retina. Modalities for imaging the retinal vessels  include color fundus (CF) photography, Fluorescein Angiography (FA), Optical Coherence Tomography Angiography (OCT-A), and scanning laser ophthalmoscope (SLO) \cite{lee2019deep}. While each modality offers complementary information, they also cause domain shift problems. 

Image alignment often relies on feature-based methods \cite{wang2015robust} for global alignment. 
These methods contain three building blocks: feature detection,  description, and  matching. Both traditional (e.g., SIFT \cite{lowe2004distinctive}, SURF \cite{bay2008speeded}, ORB \cite{rublee2011orb}) and learning-based (e.g., SuperPoint \cite{detone2018superpoint}, R2D2 \cite{revaud2019r2d2}, and SiLK \cite{gleize2023silk}) feature detection and description techniques have been developed for natural images, but they struggle with retinal images, due to illumination variations and presence of pathologies. Additionally, features are often detected along the circular perimeter of retinal images rather than at anatomically meaningful locations. Prior retina-specific models include trainable detectors for single modalities, such as GLAMpoints \cite{truong2019glampoints} and SuperRetina \cite{liu2022semi}. However, multi-modality approaches remain under-explored beyond some initial work  on domain adaptation   \cite{an2022self,sindel2022multi}. 

Another category of methods is dense feature matching techniques \cite{sun2021loftr,edstedt2023roma} that do not rely on a detector. These methods are advantageous for low-texture areas in natural images, but they often identify non-vascular regions in retinal images.

Once features are detected, traditionally, feature matching has relied on brute-force matching combined with RANSAC \cite{fischler1981random} to filter out outliers. Recent studies \cite{wang2023novel,sarlin2020superglue} have explored the use of graph-based self- and cross-attention mechanisms to train feature matching in a self-supervised manner. While some methods train a keypoint alignment framework that encompasses detection, description, and matching \cite{sindel2022multi}, separating feature detection and description from feature matching can potentially better support downstream tasks, such as identity classification \cite{liu2022semi} and prompt-based segmentation \cite{li2023promise}. 


While obtaining ground truth annotations for retinal images is challenging, sparsely annotated datasets are more feasible \cite{medal-retina,liu2022semi}. Previous methods \cite{detone2018superpoint,revaud2019r2d2,truong2019glampoints} have thus focused on self-supervised learning (SSL). Incorporating spatial features \cite{yang2024keypoint} has been shown to improve local representation learning, aiding in the identification of distinct features across modalities. Leveraging a small labeled dataset through iterative semi-supervised training \cite{liu2022semi} has also shown promise.  

\begin{figure}[t]
\label{fig:framework}
\includegraphics[width=1\columnwidth]{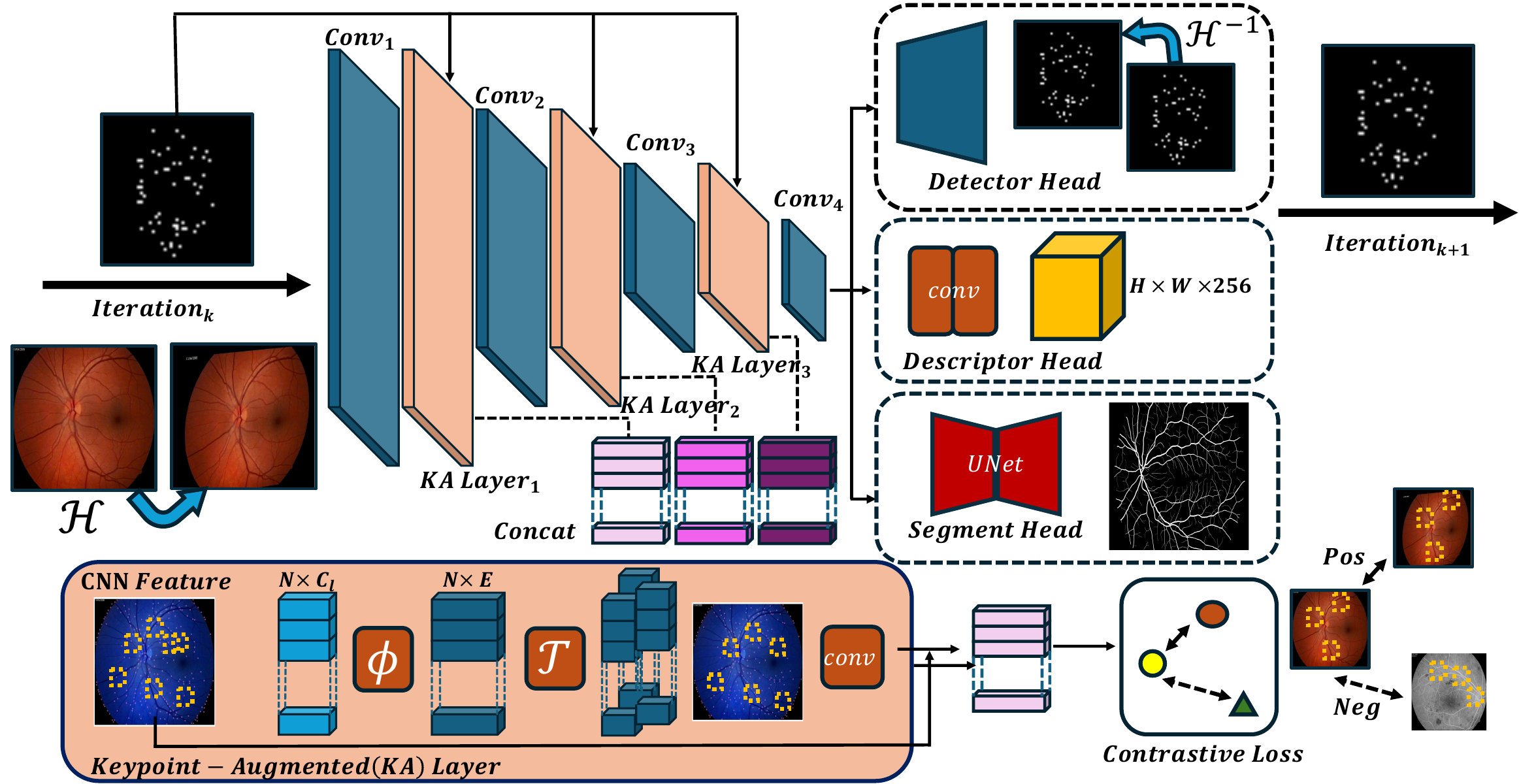}
\centering
\caption{The overall framework for retinal IPA. The bottom orange panel represents our keypoint-augmented (KA) layer, where we concat each layer result to compute the contrastive loss shown in the pink stacks. The dashed boxes represent the multi-tasking framework, with detection, description, and auxiliary segmentation tasks. In each iteration we leverage the current feature prediction to facilitate training.} 
\end{figure}

We propose retinal \underline{i}mage key\underline{p}oint \underline{a}lignment (Retinal IPA), a novel self-/semi-supervised strategy that iteratively uses the predicted keypoint candidates in training a cross-modality feature encoder (Fig.~\ref{fig:framework}).  
We evaluate our model on public and private datasets including a broad range of  modalities (fundus, FA, OCT-A, SLO).  
Our contributions are as follows:
\begin{enumerate}
    \item \textbf{Multi-tasking Integration (Sec.~\ref{sec:seg}):} By incorporating a keypoint-based segmentation branch into our training, we significantly improve consistency and robustness in feature detection across diverse transformations.
    \item \textbf{Keypoint-Augmented SSL (Sec.~\ref{sec:fusion})}: We propose a keypoint-based fusion layer companion with convolutional feature maps, capturing both short- and long-range dependencies for effective cross-modality feature encoding.
    \item \textbf{Iterative Keypoint Training (Sec.~\ref{sec:iterative}):} Through self-/semi-supervised training on a sparsely-labeled dataset, we iteratively refine the detected keypoints, progressively boosting the accuracy and reliability.
\end{enumerate}

\section{Methods}


\subsection{Datasets}
    \noindent\underline{\textbf{Training Set:}} We use the partially labeled dataset MeDAL-Retina \cite{medal-retina}, containing 208 color fundus (CF) images with human-labeled feature keypoints, with $N \in [18, 86]$ control keypoints. Additionally, it includes 1920 unlabeled CF images. We also use the OCT-500 dataset \cite{li2024octa}, featuring en-face projections of OCT/OCTA data, providing 500 2D unlabeled images. An in-house OCT-SLO mouse dataset with 228 2D unlabeled images supports multi-modal training. 

    \noindent\underline{\textbf{Test Set:}} We use two multi-modality and one single-modality datasets. The single-modality FIRE dataset \cite{hernandez2017fire} contains 134 pairs of fundus images with $2912 \times 2912$ pixels, with ground truth matching keypoints. The CF-FA dataset \cite{hajeb2012diabetic} contains 59 image pairs (Diabetic: $n=29$, Normal: $n=30$) with $720 \times 576$ pixels. Our in-house OCT-SLO human dataset contains 18 pairs of images with $1500 \times 2000$ pixels. Two annotators  manually added 8--12 keypoints to each image in the CF-FA and OCT-SLO datasets for our experiments. 

    \subsection{Background: Feature detector and descriptor overview}
    \label{sec:framework}
    
    We adopt the structure of SuperRetina \cite{liu2022semi} for our feature detector and descriptor. This model processes the input image $\mathcal{I} \in \mathbb{R}^{H \times W}$ through a convolutional encoder to produce a series of feature maps $\mathcal{F}_l$ at each downsampling level $l\in[0,3]$, where $\mathcal{F}_l \in \mathbb{R}^{\frac{H}{2^l} \times \frac{W}{2^l}\times C_l}$, and $C_l$ is the number of feature maps at level $l$. The final feature map feeds into separate decoders for the detector and the descriptor. 

\noindent
\textbf{\underline{\textit{Detector.}}}  SuperRetina poses keypoint identification as a classification task, assigning each pixel $(i, j)$ a probability $p_{i,j} \in [0,1]$ of being a keypoint. This is achieved using a U-Net architecture for the detector decoder to output a full-size probability map $\mathcal{P(I)} \in \mathbb{R}^{H \times W}$. 
We train the detector in a semi-supervised manner with labeled and unlabeled data. 

For the labeled data, let $Y$ be the keypoints associated with image $\mathcal{I}$, represented as a binary image. To compensate for the sparsity of the keypoint data, SuperRetina uses a Gaussian blur ($\sigma=0.2$ and kernel size $k=13$) on $Y$ to create a heatmap $G(Y)$. The loss function is the Dice loss between the detection probability map $\mathcal{P(I)}$ and the heatmap $G(Y)$: $L_{det-sup}=L_{Dice}(\mathcal{P(I)}, G(Y))$.   

For unlabeled data, SuperRetina assumes that detected keypoints should remain consistent across spatial transformations. 
A random homography transform $\mathcal{H}$ is used to obtain $P'=\mathcal{P}(\mathcal{H}(\mathcal{I}))$. Feature coordinates $Y'=\mathcal{C}(P')$ are extracted from the image $P'$ using a non-maxima suppresion (NMS) algorithm and thresholding at 0.5. These features are mapped back, $\mathcal{H}^{-1}(Y')$. To filter out inconsistent features, the distance between $Y$ and $\mathcal{H}^{-1}(Y')$ is thresholded at 0.5 voxels to obtain $\hat{Y}$. 
They finally apply the same Gaussian blur to obtain $G({\hat{Y}})$, which serves as supervision to compute the Dice loss, $L_{det-self} = L_{Dice}(\mathcal{P}(\mathcal{I}),G({\hat{Y}}))$.



\noindent
\textbf{\underline{\textit{Descriptor.}}} 
    The SuperRetina descriptor produces a high-dimensional vector for each keypoint, incorporating information from its neighborhood. This involves down-sampling followed by up-sampling through a transposed convolution layer, resulting in a full-size descriptor map $\mathcal{D} \in \mathbb{R}^{H \times W \times 256}$ for a 256-dimensional descriptor vector. The descriptor vectors are then L2-normalized. We use the triplet contrastive loss $L_{des}$ as defined in SuperRetina, which uses self-supervision by leveraging the assumption that descriptors should be invariant to spatial transformations while discriminative between different keypoints.

    \subsection{Contribution I: Multi-tasking keypoint-based segmentation}
    \label{sec:seg}

     We hypothesize that incorporating segmentation as an auxiliary task would enable our model to learn more domain-agnostic information to help our multi-modal performance. Given the scarcity of vessel labels for training, we use an SSL approach by again assuming invariance to transformations, for spatial transformation $\mathcal{H}$ and intensity augmentation (color jitter).

    Inspired by prior work \cite{li2022self,kirillov2023segment} that uses point prompts for segmentation, we use the predicted keypoints at each training iteration to obtain a segmentation. This is based on our observation that the keypoints even in the early stages of training tend to be vascular features. We train a U-Net \cite{ronneberger2015u} where the image and keypoints are both used as input channels.
     
    At each training iteration $k+1$, we obtain the coordinates of candidate feature points, $\mathcal{C}(\mathcal{P}_{k}(\mathcal{I}))$, by applying the NMS algorithm to the detection probability map $\mathcal{P}_{k}(\mathcal{I})$. We then create a Gaussian-blurred heatmap, $G(\mathcal{C}(\mathcal{P}_{k}(\mathcal{I})))$. This heatmap is concatenated with the original image $\mathcal{I}$ and is input to the U-Net model to produce the segmentation $\mathcal{S(I)}$. For self-supervision, we apply a homography transform, $\mathcal{H}$ to both the image and the keypoints to obtain another segmentation $\mathcal{S(H(I))}$.  The Dice loss $L_{seg}$ is then calculated between $\mathcal{S(I)}$ and $\mathcal{H}^{-1}(\mathcal{S(H(I))})$ to encourage consistency following spatial transformation.
    


    \subsection{Contribution II: Keypoint-Augmented Feature Map Level SSL}
    \label{sec:fusion}

    Inspired by Yang et al.~\cite{yang2024keypoint}, we refine the CNN encoder-decoder model to identify vascular structures across modalities with long-distance dependencies. Our proposed method leverages the feature representation by self-supervised training with iteratively adapted feature prediction. This approach diverges from shallow CNNs, which lack the capacity to capture long-distance relationships, and Vision Transformers, which are resource-intensive and demand large datasets. Unlike \cite{yang2024keypoint}, which deals with 3D volumes, we do not use a contrastive loss that focuses on spatial relationships between slices. Instead, we formulate a contrastive loss by using a homography transform $\mathcal{H}$ in a self-supervised setting.

At each iteration $k+1$, we sample the CNN encoder feature maps $\mathcal{F}_l(i, j)$ for each layer $l \in [0,2]$ at the $N_{k}$ keypoint candidates $(i, j )\in \mathcal{C}(\mathcal{P}_{k}(\mathcal{I}))$ detected in iteration $k$. 
These keypoint features are then projected to an embedding space $E \in \mathbb{R}$ via an MLP denoted as $\phi$ in Fig.~\ref{fig:framework}, followed by self-attention computation with a transformer layer ($\tau$ in Fig.~\ref{fig:framework}). 
The self-attention features are concatenated with the original convolutional features $\mathcal{F}_l$ through a dense convolution layer, serving both as input for the subsequent layer and as a skip connection for the detector decoder. 
Finally, the extracted keypoint features at each layer $l \in [0, 2]$ are concatenated and fed into a single-layer MLP, $g(\mathcal{I}) \in \mathbb{R}^ {N_k \times (3\times E)}$. 

For a training batch of $B$ images, each image $\mathcal{I}_b$ in the batch is spatially augmented by a homography transform $\mathcal{H}_b$. We obtain the keypoint features $g(\mathcal{I}_b)$ and  $g(\mathcal{H}_b(\mathcal{I}_b))$ as a positive pair of feature vectors from the same subject before and after the transform. Similarly we obtain $g(\mathcal{I}_r)$ and $g(\mathcal{H}_b(\mathcal{I}_r))$ from a random different subject in the batch ($r \in [0, B], r\neq b$) and use them as negative counterparts to $g(\mathcal{I}_b)$. We follow a similar setting as the positional contrastive learning (PCL) loss \cite{zeng2021positional}, where $cos$ represents cosine similarity and $\tau$ is the temperature term to compute the loss function $L_{ssl}$ to encourage features at corresponding locations to be similar.
    
    \begin{equation}
        L_{ssl}(g(\mathcal{I}_b), \mathcal{H}_b) = 
        - \log \frac{ e\hat{\mkern6mu} (\text{cos}(g(\mathcal{I}_b), g(\mathcal{H}_b(\mathcal{I}_b)))/\tau) / B}{\sum_{\substack{r=1 \\ r\neq b}}^{B}  [ e \hat{\mkern6mu}(\text{cos}(g(\mathcal{I}_b), g(\mathcal{I}_r)) ) +   e\hat{\mkern6mu}(\text{cos}(g(\mathcal{I}_b), g(\mathcal{H}_b(\mathcal{I}_r)))) ]/\tau} 
    \end{equation}


    The overall training objective for RetinaIPA is then 
        $\argmin_{\theta} [L_{det-sup} + L_{det-self}+ L_{des} + L_{seg}+ L_{ssl}$] where $\theta$ represents network parameters.

 \subsection{Contribution III: Iterative Keypoint Training}
    \label{sec:iterative}

   In SuperRetina \cite{liu2022semi}, the authors proposed Progressive Keypoint Expansion (PKE) to robustly and progressively add detected keypoints as labels for the supervised training, addressing the issue of partially labeled data. We have enhanced this approach to improve the model's capability by adapting the features in each iteration by using these newly detected keypoints as input to our network for self-supervised segmentation and keypoint augmentation in the next iteration (Fig.~\ref{fig:framework}). This iterative inclusion of newly detected keypoints benefits the segmentation head (Sec.~\ref{sec:seg}) by obtaining more detailed vascular maps, and allows the model to distinguish and detect new positions in the feature map (Sec.~\ref{sec:fusion}).

    \subsection{Implementation details}
    After detecting and describing features in each image, we align keypoints between image pairs. Traditional methods include nearest neighbor brute-force (nnBF) matching and RANSAC methods \cite{fischler1981random} to eliminate outliers. Learning-based SuperGlue \cite{sarlin2020superglue} and LightGlue \cite{lindenberger2023lightglue} methods employ graph-based self- and cross-attention mechanisms for enhanced matching accuracy. We directly use weights from pre-trained models for these approaches. 

    
    We rescale each image to $768 \times 768$ pixels for processing, and rescale back to original resolution for evaluation. We train our model with a batch size of $B=2$ and an initial learning rate of $1e^{-4}$. The Adam optimizer is used for a maximum of 150 epochs. Our experiments were conducted on an NVIDIA A6000 GPU (48GB memory). We use $C_0, C_1, C_2 = 64,128,128$, $E = 256$, and $\tau = 0.07$. 
       
 \begin{figure}[t]
\includegraphics[width=\columnwidth]{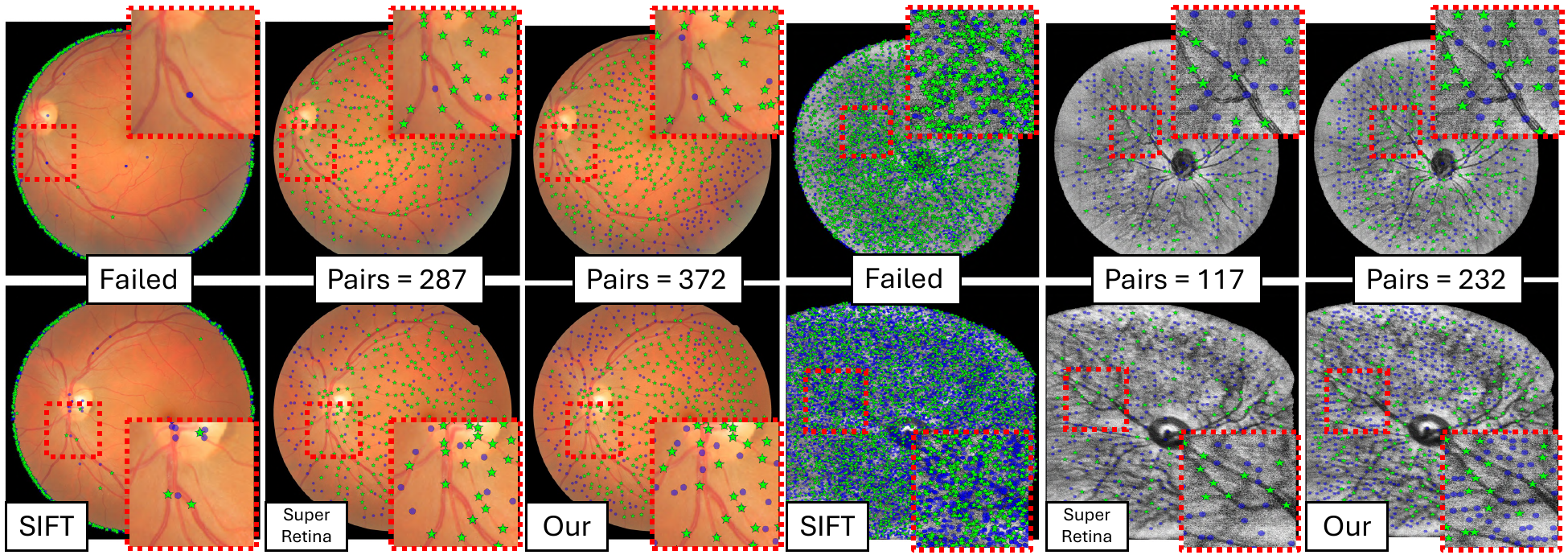}
\centering
\caption{Feature detection. First three columns: single-modality FIRE dataset. Last three columns: OCT-SLO dataset. Green stars: matched points. Blue circles: detected features. SIFT fails in both datasets. SuperRetina produces plausible results, but our model finds more matching pairs in each dataset.} 
\label{fig:quali}
\end{figure}
    
    \section{Results}

    \begin{table}[tbp]
    \centering
    \caption{Quantitative matching accuracy. The \textbf{\underline{best}}, \textbf{second} and \underline{third} performance for each metric (mMAE, mMEE, AUC) are shown with bold and underline. }
    \label{tab:quantitative_results}
    \label{123}
    \scriptsize{
    \begin{tabular}{l|c|c|c|c}
        \toprule
         \multicolumn{2}{c}{Source: MeDAL \cite{medal-retina}} & \multicolumn{3}{c}{Target (mMAE $\downarrow$ / mMEE $\downarrow$ / AUC $\uparrow$)}\\
        \hline
        Detectors  & Matching & FIRE \cite{hernandez2017fire} & CF-FA \cite{hajeb2012diabetic} & OCT-SLO \\
        \hline

        SIFT \cite{lowe2004distinctive}       & nnBF & 363 / 196 / .507 & - / - / - &  -/ - /-\\
        SuperPoint\cite{detone2018superpoint} & nnBF & 27.8 / 12.4 / .674 & 31.3 / 13.6 / .523 &   37.1/16.7/.714\\
        GLAMPoints\cite{truong2019glampoints} & nnBF & 25.3 / 11.7 / .664 & 40.2 / 22.3 / .297 & 40.0/11.2/.571 \\
        SuperRetina\cite{liu2022semi} & nnBF & 15.0 / 4.86 / \underline{.755} & 21.2 / 3.86 / .790 &  23.2/15.3/.765\\ 
        R2D2 \cite{revaud2019r2d2}      & nnBF & 19.8 / 5.78 / .683 & 62.6 / 34.9 / .568 &  33.7/17.5/.731\\
        DISK\cite{tyszkiewicz2020disk}    & nnBF & 23.2 / 5.93 / .642 & - / - / - &  43.2/27.8/.705\\
        SiLK  \cite{gleize2023silk}     & nnBF & 35.4 / 19.8 / .630 & - / - / - &  30.5/26.8/.667\\
        \hline

        - & AspanFormer\cite{chen2022aspanformer} & 15.0 / \underline{5.02} / .707 & \textbf{16.8} / \underline{2.39} / \underline{.839} &  \underline{\textbf{18.4}}/\textbf{7.92}/\underline{\textbf{.828}}\\
        - & LoFTR\cite{sun2021loftr} & 17.5 / 7.04 / .686 & 18.7 / 3.69 / .803 &  \textbf{18.5}/9.72/.744\\
        - & DKM \cite{edstedt2023dkm}& 18.5 / 9.14 / .589 & 19.1 / 2.52 / .815 &  \underline{19.7}/\underline{7.99}/\underline{.814}\\
        - & RoMa \cite{edstedt2023roma} & \underline{14.9} / \textbf{4.92} / .742 & \underline{17.0} / \textbf{2.20} /\textbf{ .848} &  20.3/\underline{\textbf{6.73}}/.799\\
        \hline

        SuperPoint & SuperGlue \cite{sarlin2020superglue} & 24.5 / 9.09 / .688 & 91.3 / 45.4 / .688 & 38.5/19.7/.711 \\
        SuperPoint & LightGlue \cite{lindenberger2023lightglue} & 20.2 / 5.43 / .705 & 87.2 / 40.8 / .716 &  43.2/25.3/.791\\
        DISK & LightGlue & 21.3 / 5.85 / .668 & 378 / 57.8 / .610 &  42.0/19.2/.720\\
        Ours & nnBF & \textbf{14.2} / 4.97 / \textbf{.761} & 18.4 / 2.99 / .808 &  20.8/14.6/.788\\
        Ours & LightGlue & \textbf{\underline{13.9}}/ \textbf{\underline{4.42}} / \textbf{\underline{.778} }& \textbf{\underline{16.3} }/\textbf{\underline{2.01}} /\textbf{\underline{ .858} }&  19.9/12.8/\textbf{.818}\\
        \hline
    \end{tabular}
    }
\end{table}


       \begin{figure}[t]
    
    \includegraphics[width=1\columnwidth]{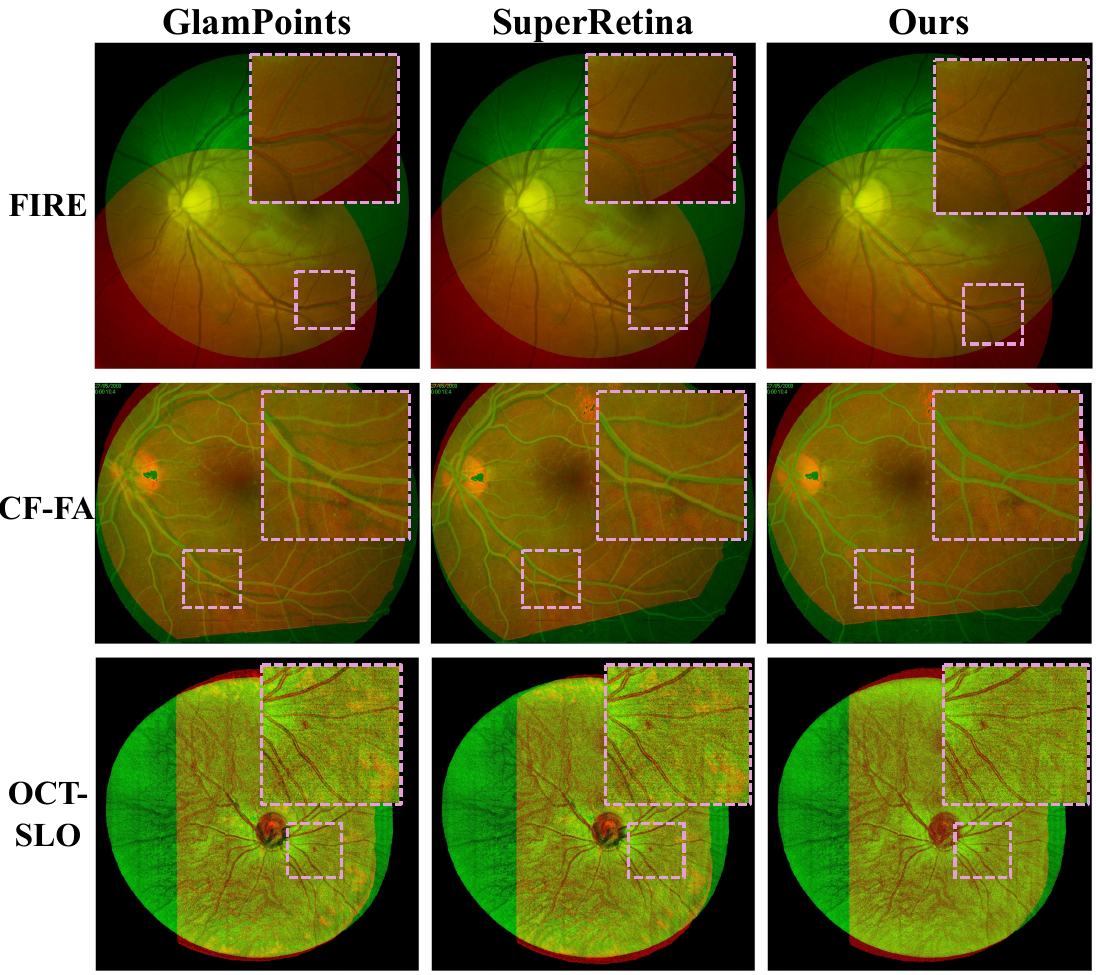}
    \centering
    \caption{Registration results. Each row is representative of a different dataset. The red channel shows the moving image after alignment ($M(I_m)$), and the green channel shows the fixed image ($I_f$). The dashed boxes provide a zoomed-in view for better visibility. We observe that our method outperforms the other two methods, which show  shadowing indicating mismatched vessels.}  
    \label{fig:matching_result}
    \end{figure}

    \noindent
    \textbf{\underline{\textit{Qualitative evaluation.}}} Fig.~\ref{fig:quali} shows the feature detection and matching results qualitatively. We observe that SuperRetina produces good results, but our method is able to find more matching pairs in each dataset. 

    \noindent
    \textbf{\underline{\textit{Alignment accuracy.}}} For each test dataset, we have sets of ground truth matching keypoints $(K_t(I_m), K_t(I_f))$ for each pair of images $(I_m, I_f)$, where $m$ and $f$ denote the moving and fixed image, respectively. We detect and align features using our model for each pair of images, and use the feature points to estimate a homography matrix $M$ that aligns $I_m$ to $I_f$ using the least median robustness algorithm \cite{rousseeuw1984least}. We then apply $M$ to the $K_t(I_m)$ and compute the L2 distance between $M(K_t(I_m))$ and $K_t(I_f)$ following prior work \cite{hernandez2017fire,liu2022semi}.
    
    In Table  \ref{tab:quantitative_results}, we report the mean of the maximum and median errors (mMAE and mMEE, respectively) across all ground truth keypoints. Additionally, we measure the area under the cumulative error curve (AUC), which corresponds to the percentage of L2 distances that fall below an error threshold set at 25 pixels \cite{liu2022semi}. We report results in each test dataset (FIRE, CF-FA, and OCT-SLO), and we note that the same criteria are applied across different resolution datasets, as we only compare models within the same dataset. We compare our methodology against state-of-the-art feature alignment methods, categorized between detector-based and detector-free methods. Note that we exclude some detector-based methods tested on the CF-FA dataset as they have performed very poorly due to the modalities being significantly different.
    
    Our methods using the LightGlue \cite{lindenberger2023lightglue} for matching outperformed all other approaches for FIRE and CF-FA datasets, and was on par for the OCT-SLO dataset. On the OCT-SLO dataset, the detector-free methods show excellent performance, even though they struggle in the FIRE and CF-FA datasets. This might suggest they are robust for handling significant discrepancies between modalities, but lack precision within a single modality.
    
    We visually compare alignment performance between $M(I_m)$ and  $I_f$ in Fig.~ \ref{fig:matching_result}. The images show significant discrepancies when applying GLAMPoints and SuperRetina, which are visible as shadowing around the vessels. Our method demonstrates superior results across all datasets, with no noticeable shadowing. 

    \noindent \underline{\textbf{\textit{Ablation Study.}}}  Our ablation study (Table \ref{tab:ablation_study_datasets}) evaluates three primary contributions of our proposed method to assess their performance both within and across modalities, using SuperRetina \cite{liu2022semi} as the baseline model. This evaluation is quantified using AUC values with an error threshold of 25 pixels. We find that adding a segmentation head (Contribution 1) enhances intra-modality performance in the FIRE dataset. In contrast, the self-supervised keypoint augmentation module (Contribution 2) improves performance in the cross-modality CF-FA and OCT-SLO datasets. Iteratively incorporating newly predicted keypoints into the network (Contribution 3) achieves even better performance in cross-modality datasets. Combined together, our methods demonstrate the best performance in all three datasets. The performance gain is significant for the multi-modal datasets.

\begin{table}[h]
    \centering
    \caption{AUC results of ablation study. Bold indicates the best results. Asterisk (*) indicates $p\leq0.05$.}
    \label{tab:ablation_study_datasets}
    \begin{tabular}{l|c|c|c}
        \toprule
        \multicolumn{4}{c}{AUC @ 25}\\

        \hline

        Target datasets & FIRE \cite{hernandez2017fire} & CF-FA \cite{hajeb2012diabetic} & OCT-SLO \\
        \hline

        Baseline \cite{liu2022semi} & 0.755 & 0.790 & 0.765\\
        Baseline+Multi-task (segmentation head) & 0.759 & 0.788 & 0.771 \\
        Baseline+Self-supervised keypoints (w/o iterative) & 0.753 & 0.791 & 0.767 \\
        Baseline+Self-supervised keypoints (w/ iterative) & 0.755 & 0.794 & 0.771 \\
        Ours (baseline+all three contributions) & \textbf{0.761} & $\textbf{0.808}^*$ & $\textbf{0.788}^*$\\
        \hline
    \end{tabular}
\end{table}

\section{Discussion}
    In this work, we introduced three novel contributions to enhance retinal image matching across multi-modality datasets. Our method integrates multi-tasking segmentation, keypoint augmentation and iterative keypoint training, significantly surpassing  methods in the natural image domain as well as those tailored for the retinal domain. Unlike existing methods that often necessitate  separate domain adaptation networks, our model can  adapt across various modalities.

\begin{credits}
\subsubsection{\ackname} This work is supported, in part, by the NIH grants R01-EY033969, R01-EY030490 and R01-EY031769.

\subsubsection{\discintname}
The authors have no competing interests to declare that are relevant to the content of this article.
\end{credits}
%
%
%
%
\clearpage
\bibliographystyle{splncs04}

\bibliography{ref.bib}

\begin{thebibliography}{10}
\providecommand{\url}[1]{\texttt{#1}}
\providecommand{\urlprefix}{URL }
\providecommand{\doi}[1]{https://doi.org/#1}

\bibitem{an2022self}
An, C., Wang, Y., Zhang, J., Nguyen, T.Q.: Self-supervised rigid registration for multimodal retinal images. IEEE Transactions on Image Processing  \textbf{31},  5733--5747 (2022)

\bibitem{bay2008speeded}
Bay, H., Ess, A., Tuytelaars, T., Van~Gool, L.: Speeded-up robust features (surf). Computer vision and image understanding  \textbf{110}(3),  346--359 (2008)

\bibitem{chen2022aspanformer}
Chen, H., Luo, Z., Zhou, L., Tian, Y., Zhen, M., Fang, T., Mckinnon, D., Tsin, Y., Quan, L.: Aspanformer: Detector-free image matching with adaptive span transformer. In: European Conference on Computer Vision. pp. 20--36. Springer (2022)

\bibitem{detone2018superpoint}
DeTone, D., Malisiewicz, T., Rabinovich, A.: Superpoint: Self-supervised interest point detection and description. In: Proceedings of the IEEE conference on computer vision and pattern recognition workshops. pp. 224--236 (2018)

\bibitem{edstedt2023dkm}
Edstedt, J., Athanasiadis, I., Wadenb{\"a}ck, M., Felsberg, M.: Dkm: Dense kernelized feature matching for geometry estimation. In: Proceedings of the IEEE/CVF Conference on Computer Vision and Pattern Recognition. pp. 17765--17775 (2023)

\bibitem{edstedt2023roma}
Edstedt, J., Sun, Q., B{\"o}kman, G., Wadenb{\"a}ck, M., Felsberg, M.: Roma: Revisiting robust losses for dense feature matching. arXiv preprint arXiv:2305.15404  (2023)

\bibitem{fischler1981random}
Fischler, M.A., Bolles, R.C.: Random sample consensus: a paradigm for model fitting with applications to image analysis and automated cartography. Communications of the ACM  \textbf{24}(6),  381--395 (1981)

\bibitem{gleize2023silk}
Gleize, P., Wang, W., Feiszli, M.: Silk--simple learned keypoints. arXiv preprint arXiv:2304.06194  (2023)

\bibitem{hajeb2012diabetic}
Hajeb Mohammad~Alipour, S., Rabbani, H., Akhlaghi, M.R.: Diabetic retinopathy grading by digital curvelet transform. Computational and mathematical methods in medicine  \textbf{2012} (2012)

\bibitem{hernandez2017fire}
Hernandez-Matas, C., Zabulis, X., Triantafyllou, A., Anyfanti, P., Douma, S., Argyros, A.A.: Fire: fundus image registration dataset. Modeling and Artificial Intelligence in Ophthalmology  \textbf{1}(4),  16--28 (2017)

\bibitem{kirillov2023segment}
Kirillov, A., Mintun, E., Ravi, N., Mao, H., Rolland, C., Gustafson, L., Xiao, T., Whitehead, S., Berg, A.C., Lo, W.Y., et~al.: Segment anything. arXiv preprint arXiv:2304.02643  (2023)

\bibitem{lee2019deep}
Lee, J.A., Liu, P., Cheng, J., Fu, H.: A deep step pattern representation for multimodal retinal image registration. In: Proceedings of the IEEE/CVF International Conference on Computer Vision. pp. 5077--5086 (2019)

\bibitem{li2022self}
Li, H., Liu, H., Hu, D., Wang, J., Johnson, H., Sherbini, O., Gavazzi, F., D’Aiello, R., Vanderver, A., Long, J., et~al.: Self-supervised test-time adaptation for medical image segmentation. In: International Workshop on Machine Learning in Clinical Neuroimaging. pp. 32--41. Springer (2022)

\bibitem{li2023promise}
Li, H., Liu, H., Hu, D., Wang, J., Oguz, I.: Promise: Prompt-driven 3d medical image segmentation using pretrained image foundation models. arXiv preprint arXiv:2310.19721  (2023)

\bibitem{li2024octa}
Li, M., Huang, K., Xu, Q., Yang, J., Zhang, Y., Ji, Z., Xie, K., Yuan, S., Liu, Q., Chen, Q.: Octa-500: a retinal dataset for optical coherence tomography angiography study. Medical Image Analysis  \textbf{93},  103092 (2024)

\bibitem{lindenberger2023lightglue}
Lindenberger, P., Sarlin, P.E., Pollefeys, M.: Lightglue: Local feature matching at light speed. arXiv preprint arXiv:2306.13643  (2023)

\bibitem{liu2022semi}
Liu, J., Li, X., Wei, Q., Xu, J., Ding, D.: Semi-supervised keypoint detector and descriptor for retinal image matching. In: European Conference on Computer Vision. pp. 593--609. Springer (2022)

\bibitem{lowe2004distinctive}
Lowe, D.G.: Distinctive image features from scale-invariant keypoints. International journal of computer vision  \textbf{60},  91--110 (2004)

\bibitem{medal-retina}
Nasser, S.A., Gupte, N., Sethi, A.: Reverse knowledge distillation: Training a large model using a small one for retinal image matching on limited data (2023), \url{https://www.dropbox.com/sh/o8q84e2eg54ay3d/AADiAkNr6bFQDoFaKeEjpYtra?dl=0}

\bibitem{revaud2019r2d2}
Revaud, J., Weinzaepfel, P., De~Souza, C., Pion, N., Csurka, G., Cabon, Y., Humenberger, M.: R2d2: repeatable and reliable detector and descriptor. arXiv preprint arXiv:1906.06195  (2019)

\bibitem{ronneberger2015u}
Ronneberger, O., Fischer, P., Brox, T.: U-net: Convolutional networks for biomedical image segmentation. In: Medical image computing and computer-assisted intervention--MICCAI 2015: 18th international conference, Munich, Germany, October 5-9, 2015, proceedings, part III 18. pp. 234--241. Springer (2015)

\bibitem{rousseeuw1984least}
Rousseeuw, P.J.: Least median of squares regression. Journal of the American statistical association  \textbf{79}(388),  871--880 (1984)

\bibitem{rublee2011orb}
Rublee, E., Rabaud, V., Konolige, K., Bradski, G.: Orb: An efficient alternative to sift or surf. In: 2011 International conference on computer vision. pp. 2564--2571. Ieee (2011)

\bibitem{sarlin2020superglue}
Sarlin, P.E., DeTone, D., Malisiewicz, T., Rabinovich, A.: Superglue: Learning feature matching with graph neural networks. In: Proceedings of the IEEE/CVF conference on computer vision and pattern recognition. pp. 4938--4947 (2020)

\bibitem{sindel2022multi}
Sindel, A., Hohberger, B., Maier, A., Christlein, V.: Multi-modal retinal image registration using a keypoint-based vessel structure aligning network. In: International Conference on Medical Image Computing and Computer-Assisted Intervention. pp. 108--118. Springer (2022)

\bibitem{sun2021loftr}
Sun, J., Shen, Z., Wang, Y., Bao, H., Zhou, X.: Loftr: Detector-free local feature matching with transformers. In: Proceedings of the IEEE/CVF conference on computer vision and pattern recognition. pp. 8922--8931 (2021)

\bibitem{truong2019glampoints}
Truong, P., Apostolopoulos, S., Mosinska, A., Stucky, S., Ciller, C., Zanet, S.D.: Glampoints: Greedily learned accurate match points. In: Proceedings of the IEEE/CVF International Conference on Computer Vision. pp. 10732--10741 (2019)

\bibitem{tyszkiewicz2020disk}
Tyszkiewicz, M., Fua, P., Trulls, E.: Disk: Learning local features with policy gradient. Advances in Neural Information Processing Systems  \textbf{33},  14254--14265 (2020)

\bibitem{wang2015robust}
Wang, G., Wang, Z., Chen, Y., Zhao, W.: Robust point matching method for multimodal retinal image registration. Biomedical Signal Processing and Control  \textbf{19},  68--76 (2015)

\bibitem{wang2023novel}
Wang, J., Li, H., Hu, D., Tao, Y.K., Oguz, I.: Novel oct mosaicking pipeline with feature-and pixel-based registration. arXiv preprint arXiv:2311.13052  (2023)

\bibitem{yang2024keypoint}
Yang, Z., Ren, M., Ding, K., Gerig, G., Wang, Y.: Keypoint-augmented self-supervised learning for medical image segmentation with limited annotation. Advances in Neural Information Processing Systems  \textbf{36} (2024)

\bibitem{zeng2021positional}
Zeng, D., Wu, Y., Hu, X., Xu, X., Yuan, H., Huang, M., Zhuang, J., Hu, J., Shi, Y.: Positional contrastive learning for volumetric medical image segmentation. In: Medical Image Computing and Computer Assisted Intervention--MICCAI 2021: 24th International Conference, Strasbourg, France, September 27--October 1, 2021, Proceedings, Part II 24. pp. 221--230. Springer (2021)

\end{thebibliography}
\end{document}